\documentclass[11pt]{article}

\usepackage[font=small,labelfont=bf]{caption}

\usepackage{enumitem}
\usepackage{graphicx}
\usepackage{xcolor}
\usepackage{blkarray}
\usepackage{amsmath,amssymb}
\usepackage{bbm}
\usepackage{mathtools}
\usepackage{pdflscape}
\usepackage{float}
\usepackage{rotating}
\usepackage{titlesec}
\usepackage{svg}
\usepackage{amsthm}

\newtheorem{theorem}{Theorem}[section]

\usepackage[most]{tcolorbox}
\tcbuselibrary{skins}

\newtcolorbox[auto counter,list inside=examplebox]{examplebox}[2][]{%
    floatplacement=h, 
    float,
    title={Box~\thetcbcounter:~#2},
    #1,
    skin=enhanced,
    titlerule=0.5pt,
    arc=0pt,
    outer arc=0pt,
    sharp corners,
}

\usepackage[
citestyle=numeric,
style=numeric,
uniquename=false,
sorting=ynt,
maxbibnames=8,
doi=false,isbn=false,url=false,eprint=false
]{biblatex}

\allowdisplaybreaks

\titleformat{\paragraph}
{\normalfont\normalsize\bfseries}{\theparagraph}{1em}{}
\titlespacing{\paragraph}
{0pt}{3.25ex plus 1ex minus .2ex}{1.5ex plus .2ex}

\DeclareMathOperator{\EX}{\mathbbm{E}}

\allowdisplaybreaks

\begin{document}

\title{Superinfection and the hypnozoite reservoir for \textit{Plasmodium vivax}: a multitype branching process approximation}

\author{Somya Mehra$^{1,2,3}$, Peter G Taylor$^{3}$}

\date{}

\maketitle

\noindent 
1: Nuffield Department of Medicine, University of Oxford, Oxford, United Kingdom \\
3: Mahidol Oxford Tropical Medicine Research Unit, Mahidol University, Bangkok, Thailand \\
3: School of Mathematics and Statistics, University of Melbourne, Parkville, Australia


\begin{abstract}
\textit{Plasmodium vivax} malaria is a mosquito-borne disease of significant public health importance. A defining feature of the within-host biology of \textit{P. vivax} is the accrual of a hypnozoite reservoir, comprising a bank of quiescent parasites in the liver that are capable of causing relapsing blood-stage infections upon activation. Superinfection, characterised by composite blood-stage infections with parasites derived from multiple mosquito inoculation or hypnozoite activation events, is another important attribute. We have previously developed a stochastic epidemic model of \textit{P. vivax} malaria, formulated as a Markov population process with countably infinitely-many types, that is adjusted for both hypnozoite accrual and blood-stage superinfection. Here, we construct a Markovian branching process with countably infinitely-many types to approximate the early stages of this epidemic model. With $P_M$ denoting the mosquito population size, we consider the limit $P_M \to \infty$ when the ratio of the mosquito and human populations is held fixed. We use a classical coupling argument to obtain a total variation bound of order $O(P_M^{2 \kappa - 1})$ that is valid until $o(P_M^\kappa)$ human-to-mosquito and mosquito-to-human transmission events have occurred, where $\kappa < 1/2$ is an arbitrary constant. We characterise the probability of global disease extinction under the branching process to approximate the probability of elimination, as opposed to sustained endemic transmission, when the epidemic model is initialised with low-level human and/or mosquito infection. We apply our model to two scenarios of epidemiological interest, namely the re-introduction of \textit{P. vivax} malaria in a region where elimination has previously been achieved; and a mass drug administration campaign with population-wide depletion of the hypnozoite reservoir.
\end{abstract}

\textit{Plasmodium vivax} is a mosquito-borne parasite that persists as a significant cause of malarial morbidity across endemic regions in Asia, the Americas and the Western Pacific. Transmission to human hosts involves parasite inoculation through the bite of an infected anopheline mosquito. Inoculated parasites may either undergo immediate development, manifesting as a primary blood-stage infection within a typical period of nine days \parencite{mikolajczak2015plasmodium}, or take a quiescent form in the liver known as a hypnozoite. The accrual of a hypnozoite bank over sequential inoculation events is a key attribute of the epidemiology of \textit{P. vivax}, with the activation of an individual hypnozoite sufficient to cause an additional blood-stage infection termed a relapse \parencite{imwong2012first}. While putative mechanisms of hypnozoite activation are not known, our previous work has shown that a simple within-host model, formulated as an open network of infinite server queues where each hypnozoite is governed by an independent process and activates spontaneously at a constant rate \parencite{white2014modelling, mehra2022hypnozoite}, can largely reproduce observed temporal patterns of \textit{P. vivax} blood-stage infection \parencite{Mehra2024b}.  

In human hosts, blood-stage infections may comprise `broods' of parasites derived from multiple primary infections or hypnozoite activation events, a phenomenon known as superinfection. Akin to \parencite{mehra2024superinfection}, we characterise \textit{P. vivax} blood-stage infections by the multiplicity of broods (MOB), defined to be the unique number of hypnozoite activation events and primary infections contributing to the circulating blood-stage parasite biomass. Classical models of malarial superinfection (formulated for the non-relapsing malarial species \textit{Plasmodium falciparum}) assume that each brood is governed by an independent stochastic process, and subject to natural clearance at a constant rate \parencite{bailey1957, dietz1974malaria}. Since transmission to mosquito vectors involves the ingestion of circulating parasites when a mosquito takes a bloodmeal from a blood-stage infected human, superinfection may be an important determinant of onward transmission.

In \textcite{mehra2024superinfection}, we propose a stochastic epidemic model of \textit{P. vivax} malaria allowing for hypnozoite accrual and superinfection. We formulate this model as a Markov population process with countably infinitely-many types, where human hosts are stratified by both hypnozoite burdens and MOB. A functional law of large numbers, comprising a deterministic compartment model defined by an infinite-dimensional system of ordinary differential equations (ODEs), is obtained in the infinite population limit $P_M \to \infty$ (with the ratio of the human to mosuito population held fixed) and analysed. Here, we seek to construct a branching process to approximate the early stages of the epidemic model of \parencite{mehra2024superinfection}, initialised with low-level human and mosquito infection. Our primary motivation is to characterise the probability of disease elimination, as opposed to sustained endemic transmission, following a period of interrupted transmission or targeted control (for example, a mass drug administration campaign) during which the the reservoir of infection in both humans and mosquitoes is substantially depleted, or when disease is reintroduced in a setting in which elimination had previously been achieved.

The approximating branching process, which is Markovian and possesses countably infinitely-many types, is constructed in Section \ref{sec::model_structure}. The underlying intuition is two-fold. In the early stages of an epidemic with a low rate of mosquito-to-human transmission, co-circulating broods derived from the \textit{same} mosquito inoculum constitute the dominant source of superinfection; and the depletion of susceptible mosquitoes is negligibly low. In Section \ref{sec::coupling}, we use an argument analogous to that of \textcite{ball1995strong} to couple the epidemic model of \parencite{mehra2024superinfection} to the approximating branching process. In the infinite population limit $P_M \to \infty$ (likewise with the ratio of the mosquito to human population held fixed), we obtain a total variation bound of order $O(P_M^{2 \kappa - 1})$ that is valid until $o(P_M^\kappa)$ human-to-mosquito and mosquito-to-human transmission events have occurred, where $\kappa < 1/2$ is an arbitrary constant (Theorem \ref{theorem::superinf_branching_approx}a). Along sample paths where disease extinction occurs in the approximating branching process, we additionally establish strong convergence of the epidemic model in the limit $P_M \to \infty$ (Theorem \ref{theorem::superinf_branching_approx}b). To approximate the probability of disease elimination vs sustained endemic transmission when the epidemic process is initialised with low-level human/mosquito infection, we characterise the probability of global extinction in the branching process (Section \ref{sec::disease_extinction}, Theorem \ref{theorem::branching_process_extinction}). Illustrative results are presented in Section \ref{sec::illustrative_results} for two scenarios of epidemiological interest: the probability of re-instated transmission when a recently-inoculated individual enters a setting where \textit{P. vivax} has previously been eliminated; and the probability of elimination following an idealised mass drug administration campaign which leads to substantial depletion of the population-wide hypnozoite burden, and perfect clearance of pre-existing blood-stage and mosquito infections (the latter due to an extended period of blood-stage prophylactic protection which succeeds in temporarily interrupting transmission). Concluding remarks are provided in Section \ref{sec::discussions}.


\section{Model structure} \label{sec::model_structure}

Adopting the notation of \textcite{mehra2024superinfection}, we interpret 
\begin{itemize}
    \item $h_{i,j}(t)$, $i,j \in \mathbbm{Z}_{\geq 0}$ to be the number of humans with hypnozoite burden $i$ and MOB $j$;
    \item $m_i(t)$, $i \in \{0, 1 \}$ to be the number of uninfected and infected mosquitoes respectively.
\end{itemize}

For the epidemic model, we set
\begin{align*}
    \mathbf{h}(t) &= (h_{0,0}(t), h_{0,1}(t), h_{1,0}(t), h_{0, 2}(t), h_{1, 1,}(t), h_{2,0}(t), \dots),
\end{align*}
whereby $h_{i,j}(t)$ occupies the $\big( \frac{1}{2}(i+j+1)(i+j) + i +1\big)^{\text{th}}$ entry of $\mathbf{h}(t)$, and denote by $\mathbf{e_{i,j}}$ the corresponding (unit) coordinate vector in $\mathbbm{R}^\mathbf{N}$. We additionally set $\mathbf{m}(t) = (m_0(t), m_1(t))$.

For the branching process approximation, we remove the component with zero human infection to yield
\begin{align*}
    \mathbf{h^*}(t) &= (h_{0,1}(t), h_{1,0}(t), h_{0, 2}(t), h_{1, 1,}(t), h_{2,0}(t), \dots),
\end{align*}
with the coordinate vectors $\mathbf{e^*_{i,j}}$ defined similarly, and denote by $m_1^*(t)$ the component of infected mosquitoes.

\subsection{An epidemic model with superinfection and hypnozoite accrual}
Here, we consider the Markov population process constructed in \textcite{mehra2024superinfection} to describe \textit{P. vivax} dynamics in a closed population of $P_M$ mosquitoes and $P_H$ humans, whilst accommodating (blood-stage) superinfection and hypnozoite accrual. For clarity, we have reparametrised the model to ignore the contribution of hypnozoite death by pre-emptively thinning incoming inocula to retain only hypnozoites that are destined to activate; this was adopted in \parencite{Mehra2024b}, because hypnozoite death is unobservable given data pertaining to blood-stage recurrences only (which constitutes the bulk of available epidemiological data).

The underlying assumptions of the model, as detailed in \textcite{mehra2024superinfection}, are:
\begin{itemize}
    \item Contact between each mosquito and the human population occurs at constant rate $\beta$.
    \item An uninfected mosquito acquires infection with probability $q$ when it bites a human with MOB$\geq$1.
    \item A mosquito-to-human transmission event occurs with probability $p$ when an infected mosquito bites a human and establishes, irrespective of the prior human infection status:
    \begin{itemize}
        \item[(i)] an additional primary blood-stage infection; and
        \item [(ii)] a geometrically-distributed hypnozoite inoculum of mean size $\nu$ and state space $\mathbbm{Z}_{\geq 0}$, with i.i.d. batch sizes across bites.
    \end{itemize}
    \item The time until a hypnozoite activates (to give rise to a blood-stage relapse) is exponentially distributed with rate parameter $\eta$. The activation times of different hypnozoites are independent.
    \item The time until a blood-stage infection (primary or relapse) is cleared is exponentially distributed with rate parameter $\gamma$. The clearance times of different blood-stage infections are independent.
    \item To accommodate mosquito death whilst fixing the mosquito population size, infected mosquitoes are subject to `replacement' by an uninfected counterpart at constant rate $g$.
\end{itemize}

We index the epidemic process by the size of the mosquito population $P_M$ and the ratio of the mosquito and human population sizes $r = P_M/P_H$. On the state space
\begin{align*}
    \chi_{r, P_M} = \big\{ (\mathbf{h}, \mathbf{m}) \in [0, P_H]^{\mathbbm{N}} \times [0, P_M]^2: |\mathbf{h}|_1 = P_M/r, |\mathbf{m}|_1=P_M \big\},
\end{align*}
this yields a Markov population process $Y^{P_M}_r$ with transition rates
\begin{align}
    q_{ ( \mathbf{h}, \mathbf{m}), ( \mathbf{h} -  \mathbf{e_{i, j}} + \mathbf{e_{i, j-1}}, \mathbf{m})} = \gamma j h_{i, j} & \qquad (i,j) \in {\mathbb Z^2_{\geq 0}} \label{eq::blood_clearance_epidemic}\\
    q_{ ( \mathbf{h}, \mathbf{m}), ( \mathbf{h} -  \mathbf{e_{i, j}} + \mathbf{e_{i-1, j+1}}, \mathbf{m})} = \eta i h_{i, j} & \qquad (i,j) \in {\mathbb Z^2_{\geq 0}} \\
    q_{ ( \mathbf{h}, \mathbf{m}), ( \mathbf{h} -  \mathbf{e_{i, j}} + \mathbf{e_{i+\ell, j+1}}, \mathbf{m})} = \frac{\beta r p \nu^\ell}{(\nu + 1)^{\ell+1}}  \frac{m_1}{P_M} h_{i, j} & \qquad (i,j,\ell) \in {\mathbb Z^3_{\geq 0}} \label{eq::m_to_h_epidemic} \\
    q_{( \mathbf{h}, \mathbf{m}), ( \mathbf{h}, \mathbf{m} - \mathbf{e_1} + \mathbf{e_{0}})} = g m_1 & \label{eq::mosquito_death_epidemic} \\
    q_{( \mathbf{h}, \mathbf{m}), ( \mathbf{h}, \mathbf{m} - \mathbf{e_0} + \mathbf{e_1})}  = \beta q  r \Big( \sum^\infty_{i=0} \sum^\infty_{j=1} \frac{h_{i,j}}{ P_M} \Big) m_0. \label{eq::h_to_m_epidemic} 
\end{align}

\textbf{\emph{Remark}}: There is a typographical mistake in Equation (10) of \textcite{mehra2024superinfection}, pertaining to the transition rate for mosquito-to-human transmission. The constant $r$ in the numerator is missing. This is corrected in Equation (\ref{eq::m_to_h_epidemic}) above. This missing constant carries through to Equation (16) of \parencite{mehra2024superinfection} and persists in Appendix A where we verify conditions for the functional law of large numbers to hold; however, with appropriate modification, all results stated in \parencite{mehra2024superinfection} carry through.

\subsection{The multitype branching process approximation} \label{sec::branching_approximation}

To approximate the early stages of the epidemic process $Y_r^{P_M}$, we construct a Markovian branching process with countably infinitely-many types. Accounting for batch structure within mosquito inocula is imperative: hypnozoites established through the same infective bite can activate in quick succession, or in close temporal proximity to the accompanying primary infection, to give rise to blood-stage superinfection in the early stages of the epidemic process. Assuming a homogeneous transmission potential for each hypnozoite would yield a much simpler branching process with finitely-many types; but would serve as a poor approximation if the time scale of blood-stage infection was sufficiently long to permit superinfection derived from the same mosquito batch.

While the original epidemic process is constructed on the state space $\chi_{r, P_M}$, we construct the approximating process on the modified state space 
\begin{align*}
    \chi^* = \big\{ (\mathbf{h^*}, m_1^*) \in \mathbbm{Z}_{\geq 0} ^{\mathbbm{N}} \times \mathbbm{Z}_{\geq 0} \big\},
\end{align*}
in which the respective components of uninfected humans and mosquitoes are removed, and the number of infected humans and mosquitoes is unbounded.

On the state space $\chi^*$, we formulate a Markovian multitype branching process $X_r$ with transition rates
\begin{align}
    q_{ ( \mathbf{h^*}, m_1^*), ( \mathbf{h^*} -  \mathbf{e^*_{i, j}} + \mathbf{e^*_{i, j-1}}, m_1^*)} = \gamma j h^*_{i, j}, & \qquad (i,j) \in {\mathbb Z^2_{\geq 0}} \setminus \{(0,0),(0,1)\} \label{eq::human_clearance_branching_1} \\
    q_{ ( \mathbf{h^*}, m_1^*), ( \mathbf{h^*} -  \mathbf{e^*_{0, 1}}, m_1^*)} = \gamma h^*_{0, 1}, & \label{eq::human_clearance_branching} \\
    q_{ ( \mathbf{h^*}, m_1^*), ( \mathbf{h^*} -  \mathbf{e^*_{i, j}} + \mathbf{e^*_{i-1, j+1}}, m_1^*)} = \eta i h^*_{i, j}, & \qquad (i,j) \in {\mathbb Z^2_{\geq 0}} \setminus \{(0,0)\}  \\
    q_{ ( \mathbf{h^*}, m_1^*), ( \mathbf{h^*}  + \mathbf{e^*_{\ell, 1}}, m_1^*)} = \frac{\beta p \nu^\ell}{(\nu + 1)^{\ell+1}}  m_1^* & \qquad \ell \in \mathbb Z_{\geq 0} \label{eq::m_to_h_branching}   \\
    q_{( \mathbf{h^*}, m_1^*), ( \mathbf{h^*}, m_1^* - 1)} = g m_1^* & \label{eq::mosquito_death_branching} \\
    q_{( \mathbf{h^*}, m_1^*), ( \mathbf{h}^*, m_1^*+1)}  = \beta q r \Big( \sum^\infty_{i=0} \sum^\infty_{j=1} h^*_{i,j} \Big). \label{eq::h_to_m_branching}
\end{align}

The branching process $X_r$ counts only infected mosquitoes and humans with at least one blood-stage infection or hypnozoite. Upon death, infected mosquitoes are removed from the branching process $X_r$ (Equation (\ref{eq::mosquito_death_branching})); in contrast, infected mosquitoes are replaced by uninfected progeny upon ``death'' in the epidemic process $Y_r^{P_M}$ (Equation (\ref{eq::mosquito_death_epidemic})) to maintain a constant mosquito population size $P_M$. Similarly, a human with zero hypnozoites is removed from the branching process $X_r$ upon clearance of their final blood-stage infection (Equation (\ref{eq::human_clearance_branching})); an analogous human in the epidemic process would instead enter the uninfected state (Equation (\ref{eq::blood_clearance_epidemic})). We incorporate the parameter $r$ in transition rate (\ref{eq::h_to_m_branching}) of the branching process approximation for comparability to the epidemic model: this scales the magnitude of human-to-mosquito transmission (relative to mosquito-to-human transmission) in the branching process $X_r$ in a way that adjusts for differences in the (fixed) human versus mosquito population size in the epidemic process $Y_r^{P_M}$.

The branching process $X_r$ differs from the epidemic process $Y_r^{P_M}$ on two grounds. 
First, 
each mosquito-to-human transmission event gives rise to a \textit{new} individual harbouring $\ell \geq 0$ hypnozoites and one primary infection (Equation (\ref{eq::m_to_h_branching})). This means that blood-stage superinfection in an individual can  occur only when a hypnozoite activates. Furthermore, this hypnozoite must necessarily come from the same inoculation as the pre-existing blood-stage infection, whether it be a primary infection or relapse. 
Consequently, under the branching process $X_r$, $h^*_{i,j}(t)$ can equivalently be interpreted as the number of distinct mosquito inocula to which $i$ hypnozoites and $j$ blood-stage infections can be attributed at time $t$. In contrast, in the epidemic process $Y_r^{P_M}$, a previously-infected human may receive an additional primary infection and hypnozoite batch upon receiving an infective mosquito bite (Equation (\ref{eq::m_to_h_epidemic})). Therefore, $h_{i,j}(t)$ under the epidemic process $Y_r^{P_M}$ corresponds to the number of humans with $i$ hypnozoites and $j$ blood-stage infections at time $t$, where it is possible for humans to concurrently harbour hypnozoites and blood-stage infections derived from multiple mosquito bites. In the branching process model $X_r$, the rate of human-to-mosquito transmission thus scales linearly with the number of distinct mosquito inocula, summed over the human population, from which circulating blood-stage infections are derived (Equation (\ref{eq::h_to_m_branching})), as opposed to the number of blood-stage infected humans in the epidemic process $Y_r^{P_M}$ (Equation (\ref{eq::h_to_m_epidemic})). In the early stages of an epidemic with relatively infrequent mosquito-to-human transmission, the dominant source of superinfection would likely be overlapping relapses and primary infections derived from the same mosquito bite, rendering this approximation reasonable.

Secondly, in the branching process approximation, we do not adjust for the depletion of susceptible mosquitoes; in effect, we assume the availability of an infinitely large pool of mosquito vectors. The rate of human-to-mosquito transmission in the epidemic process $Y_r^{P_M}$ (Equation (\ref{eq::h_to_m_epidemic})) is scaled down by the proportion of non-infected mosquitoes $m_0/P_M$ remaining in the population; this adjustment is not applied to the rate of human-to-mosquito transmission in the branching process $X_r$ (Equation (\ref{eq::h_to_m_branching})). Likewise, this assumption is plausible in the early stages of an epidemic when the prevalence of infection in the mosquito population is limited.

\section{Coupling  the epidemic model to the approximating branching process} \label{sec::coupling}

We use an argument analogous to that of \textcite{ball1995strong} to couple the epidemic model $Y^{P_M}_r$ to the Markovian branching process $X_r$. When taking scaling limits, we fix the ratio of the mosquito to human population size $r = P_M/P_H$ and let $P_M \to \infty$. In Theorem \ref{theorem::superinf_branching_approx}, we establish a total variation bound of order $O(P_M^{2\kappa-1})$ in the limit $P_M \to \infty$ for paths comprising $o(P_M^\kappa)$ human-to-mosquito and mosquito-to-human transmission events, where $\kappa < 1/2$ is an arbitrary constant. We additionally establish strong convergence in the limit $P_M \to \infty$ along sample paths where disease extinction occurs in the branching process.

Akin to \parencite{barbour2010coupling}, we express Theorem \ref{theorem::superinf_branching_approx} with respect to the Borel $\sigma$-algebra $\mathcal{H}_n$ generated by paths in $\chi^*$ with up to $n$ transmission events, that are initialised with some state $\xi(0) \in \chi^*$. We define this as follows. For ease of construction, we first modify the processes $X_r$ and $Y^{P_M}_r$ to introduce `jumps' at unit rate from the otherwise absorbing state $\mathbf{0}$ to itself \parencite{barbour2010coupling}. We then formulate paths with countably infinitely-many transitions in the state space $\chi^*$ as infinite-dimensional vectors of random variables, each taking values in $\chi^* \times \mathbbm{R}$. We define the Borel $\sigma$-algebra $\mathcal{B} (\chi^* \times \mathbbm{R})$ on $\chi^* \times \mathbbm{R}$ as that generated by the discrete topology on $\chi^*$ and the open sets in $\mathbbm{R}$. The Borel $\sigma$-algebra generated by paths in $\chi^*$ with countably infinitely-many transitions is then taken to be the product $\sigma$-algebra $\mathcal{B} ((\chi^* \times \mathbbm{R})^\infty) := (\mathcal{B} (\chi^* \times \mathbbm{R}))^\infty$ (Lemma 1.2 of \parencite{kallenberg2021foundations}). 

To restrict our attention to paths with a given number of transmission events, we consider the set of jumps $\mathcal{J} := \{ (\mathbf{e}_{k,1}, 0), \, k \in \mathbbm{Z}_{\geq 0} \} \cup \{ (\mathbf{0}, 1) \}$ associated with either mosquito-to-human or human-to-mosquito transmission. We then define a function 
\begin{align*}
    a: (\mathcal{\chi}^* \times \mathbbm{R})^\infty \to \mathbbm{Z}, \, a \bigg(\prod^\infty_{i=0} (\mathbf{x}_i, t_i) \bigg) = \sum^\infty_{i=0} \mathbbm{1} \{ \mathbf{x}_{i+1} - \mathbf{x}_i \in \mathcal{J} \} 
\end{align*}
to return the (possibly infinite) number of transmission events present in a path with countably infinitely-many transitions. The Borel $\sigma$-algebra on the set of paths initialised in state $\xi^{(0)}$ with up to $n$ transmission events $\mathcal{H}_n := \mathcal{B}(\{ \mathbf{y} \in (\mathcal{\chi}^* \times \mathbbm{R})^\infty | a (\mathbf{y}) \leq n, \, y_0 = (\xi^{(0)}, 0) \})$ is then given by the subspace topology induced by $\mathcal{B} ((\chi^* \times \mathbbm{R})^\infty)$.

For notational simplicity, for $\xi \in \chi^*$ or $\xi \in \chi_{r, P_M}$, we denote by $b(\mathbf{\xi}) = \sum^\infty_{i=0} \sum^\infty_{j=1} h_{i,j}$ the number of blood-stage infected humans, and by $m(\mathbf{\xi}) = m_1$ the number of infected mosquitoes.

\begin{theorem} \label{theorem::superinf_branching_approx}
Suppose that $\xi^{(0)} \in \chi^*$. 
\begin{enumerate}
    \item[(a)] Let $\mathcal{H}_n$ be the Borel $\sigma$-algebra generated by paths in $\chi^*$ initialised in the state $\xi^{(0)} \in \chi^*$ with at most $n$ transmission events (human-to-mosquito and mosquito-to-human combined). Denote by $Y_r^{P_M *}$ the process $Y_r^{P_M}$ with the respective components for uninfected humans and uninfected mosquitoes removed. For any $W \in \mathcal{H}_n$,
    \begin{align*}
        \big| \mathbbm{P}[X_r \in W ] - \mathbbm{P}[X_r^{P_M *} \in W ] \big| \leq \frac{(n+b(\xi^{(0)}) +m(\xi^{(0)}) )^2}{2 P_M} (1 + r).
    \end{align*}
    \item [(b)] Let $(\Omega, \mathcal{F}, \mathbbm{P})$ be a probability space on which $X_r(\cdot)$ and the sequence $\{ Y_r^{P_M}(\cdot), P_M \geq 1\}$ is defined. Denote by $A:= \{ \omega \in \Omega: \lim_{t \to \infty} X_r(t, \omega) = 0 \}$ the set of sample paths of the branching process $X_r$ which culminate in disease extinction ($\mathbbm{P}(A)=1$ if $X_r$ is subcritical; $\mathbbm{P}(A)<1$ if $X_r$ is supercritical). Then in the limit $P_M \to \infty$
    \begin{align*}
        \sup_{0 \leq t < \infty} \big| Y_r^{P_M}(t, \omega) - X_r(t, \omega) \big|_1 \to 0 \text{ for } \mathbbm{P}\text{-almost all } \omega \in A.
    \end{align*}
\end{enumerate}
\end{theorem} 

\begin{proof}
Denote by $(\Omega_1, \mathcal{F}_1, \mathbbm{P}_1)$ the probability space on which the branching process $X_r(\cdot)$ is defined. Akin to \textcite{ball1995strong}, we draw on the waiting time process associated with the birthday problem to couple the epidemic model with superinfection to a branching process. Let $\mathcal{M}^{(P_M, r)}$ be the number of draws with replacement from two independent urns, containing $P_M$ and $P_M/r  = P_H$ balls respectively, until a previously-drawn ball is drawn from either urn. Then
\begin{align*}
    P( \mathcal{M}^{(P_M, r)} > n) = \prod^{n-1}_{m=0} \Big( 1 - \frac{m}{P_M} \Big) \Big( 1 - \frac{r m}{ P_M} \Big).
\end{align*}
It is straightforward to establish weak convergence of the sequence
\begin{align*}
    P_M^{-\frac{1}{2}} \mathcal{M}^{(P_M, r)} \implies \mathcal{M}
\end{align*}
where $\mathcal{M}$ has density
\begin{align*}
    f(x) = ( 1 + r ) x e^{-\frac{1}{2} (1+r) x^2}
\end{align*}
(see Exercise 3.1.1 of \parencite{durrett2019probability}).

By Skorokhod's representation theorem \parencite{billingsley2013convergence}, we can thus construct a probability space such that
\begin{align*}
    P_M^{-\frac{1}{2}} \mathcal{M}^{(P_M, r)} \to \mathcal{M} \text{ almost surely as } P_M \to \infty.
\end{align*}

Here, we consider an extended probability space $(\Omega_2, \mathcal{F}_2, \mathbbm{P}_2)$ which carries, for each $P_M$ and $s \in \{h, m\}$, sequences of random variables $\psi_1^{(P_M,s)}, \psi_2^{(P_M,s)}, \dots$ such that
\begin{align*}
    \psi_i^{(P_M,h)} \, \overset{\text{i.i.d}}{\sim} & \; \; \mathcal{U} \big\{1, P_M/r \big\} \qquad \qquad \psi_i^{(P_M,m)}  \overset{\text{i.i.d}}{\sim} \, \mathcal{U} \big\{ 1, P_M \big\}
\end{align*}
with
\begin{align*}
    \mathcal{M}^{(P_M, r)} = & \inf \Big\{ k: k > 1, \, \psi^{(P_M,h)}_k \in \big\{ \psi^{(P_M,h)}_1, \dots, \psi^{(P_M,h)}_{k-1} \big\} \text{ or } \psi^{(P_M,m)}_k \in \big\{ \psi^{(P_M,m)}_1, \dots, \psi^{(P_M,m)}_{k-1} \big\} \Big\},
\end{align*}
as well as a sequence of random variables $u_1, u_2, \dots$ such that $u_i \overset{\text{i.i.d.}}{\sim} \mathcal{U}[0, 1]$. 

On the probability space $(\Omega_1, \mathcal{F}_1, \mathbbm{P}_1)$, we characterise the branching process by associating with each `birth':
\begin{itemize}
    \item A type $S$ (humans with $i$ hypnozoites and MOI $j$ are designated type $H_{i,j}$; infected mosquitoes are designated type $M$);
    \item A lifetime $I$, after which the individual is either removed from the branching process (in the case $S=M$ or $S=H_{0,1}$) or transitions to another type (i.e., $S=H_{i, j} \to H_{i-1, j+1}$ for $i \geq 1$ or $S=H_{i, j} \to H_{i, j-1}$ for $j>1$ or $i>0, j=1$);
    \item A marked point process $\zeta$ over the interval $[0, I]$ governing the timing of onward human-to-mosquito transmission events (in the case $S=H_{i,j}$, $j \geq 1$) or mosquito-to-human transmission events (in the case $S=M)$, and the type of the resultant child particle (in the case $S=H_{i,j}$, the child is necessarily of type $M$; in the case $S=M$, the child is of type $H_{\ell, 1}$, $\ell \geq 0$).
\end{itemize}
The tuple $(S, I, \zeta)$ is collectively termed the `life history' of the individual.

On the product space $(\Omega, \mathcal{F}, \mathbbm{P}) = (\Omega_1 \times \Omega_2, \mathcal{F}_1 \times \mathcal{F}_2, \mathbbm{P}_1 \times \mathbbm{P}_2)$, we extend the branching process to assign a label $L$ (ranging from $\{1, \dots, P_M/r  \}$ in the case $S=H_{i,j}$ and $\{1, \dots, P_M\}$ in the case $S=M$) to each individual. The initial $b(\xi^{(0)})$ infected humans are assigned labels $\psi_{1}^{(P_M, h)}, \dots \psi_{b(\xi^{(0)})}^{(P_M, h)}$; while the initial $m(\xi^{(0)})$ infected mosquitoes are ascribed labels $\psi_{b(\xi^{(0)}) + 1}^{(P_M, m)}, \dots \psi_{b(\xi^{(0)}) + m(\xi^{(0)})}^{(P_M, m)}$. For `births' after time zero, labels are assigned as follows.  In the case of a transition of type $S=H_{i, j} \to H_{i-1, j+1}$ for $i \geq 1$ or $S=H_{i, j} \to H_{i, j-1}$ for $j>1$ or $i>0, j=1$, the label is inherited from the parent particle. Otherwise, at the $k^{\text{th}}$ transmission event in the process, we assign the label $L=\psi_{k + b(\xi^{(0)}) + m(\xi^{(0)})}^{(P_M, h)}$ if the daughter particle is of type $S = H_{\ell, 1}$, $\ell\geq0$ (mosquito-to-human transmission) or $L=\psi_{k + b(\xi^{(0)}) + m(\xi^{(0)})}^{(P_M, m)}$ if the daughter particle is of type $S=M$ (human-to-mosquito transmission). 

For each $P_M$, we couple a realisation of the epidemic process to the branching process on $(\Omega, \mathcal{F}, \mathbbm{P})$. We first thin the branching process at human-to-mosquito transmission events, or equivalently the births of type $M$ particles. Let $0 \leq \tau_1 \leq \tau_2 \dots$ denote the birth times of particles in the branching process. We count the number of human-to-mosquito and mosquito-to-human transmission events $A(k)$ that have occurred up to $\tau_k$ in the branching process (present birth included). If a particle of type $S \neq M$ is generated at time $\tau_k$ in the branching process, then a particle with the same life history is added to the thinned process. Otherwise, we compute the \textit{cardinality} $c_k$ of the set of labels of blood-stage infected humans (of type $H_{i,j}$, $j \geq 1$) in the thinned process; in addition to the number $d_k$ of blood-stage infected humans in the branching process. If $u_k \leq c_k/d_k$ \emph{and} no pre-existing individuals of type $M$ already carry the label $\chi_{A(k) + b(\xi^{(0)}) + m(\xi^{(0)})}^{(P_H, m)}$ at time $\tau_k$, then a particle of type $M$ is born in the thinned process and is ascribed an identical life history to the corresponding particle in the branching process. Otherwise, the particle of type $M$ born at time $\tau_k$ in the branching process, and its descendents thereafter, do not appear in the thinned process. This adjusts both for the double-counting of the human-to-mosquito transmission potential of superinfections derived from different mosquito inocula (Figure \ref{sec::branching_approximation}), and the depletion of susceptible mosquitoes as disease progresses (discussed at length in Section \ref{sec::branching_approximation}).

\begin{figure}[h]
    \centering
    \includegraphics[width=\textwidth]{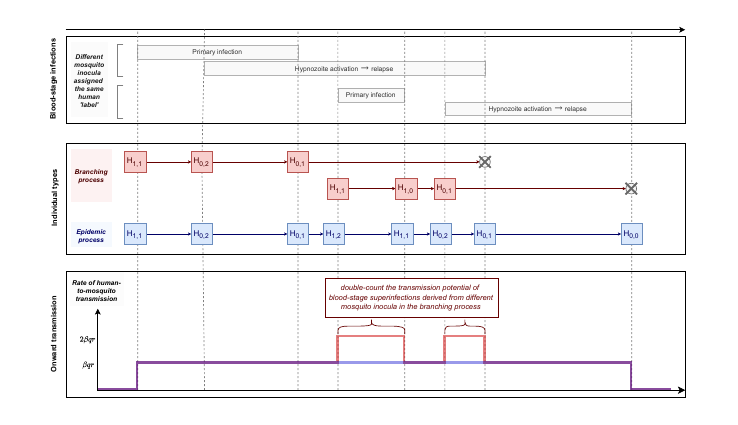}
    \caption{Schematic illustrating the double-counting of the human-to-mosquito transmission potential of superinfections derived from different mosquito inocula under the branching process $X_r$ relative to the epidemic process $Y_r^{P_M}.$}
    \label{fig:double_counting}
\end{figure}

We then construct a map between the thinned process and a realisation of the epidemic process. For each $x=1, \dots, P_M$, mosquito $x$ is said to be infected at time $t$ in the epidemic process if a particle of type $M$ with label $x$ is present in the thinned process at time $t$. For each $y=1, \dots, P_M/r$, to determine the state of human $y$ at time $t$ in the epidemic process, we first construct a multiset $S^y_t =\{ H_{i_1, j_1}, \dots, H_{i_n, j_n} \}$ by recording the types of humans with label $y$ in the epidemic process at time $t$. Human $y$ is then said to harbour $i_1 + \dots + i_n$ hypnozoites and $j_1 + \dots + j_n$ blood-stage infections at time $t$ in the epidemic model.

We then make the observation that the branching process and the $P_M^\text{th}$ thinned process are necessarily indistinguishable until the $\big( \mathcal{M}^{(P_M, r)} - b(\xi^{(0)}) - m(\xi^{(0)}) \big)^\text{th}$ human-to-mosquito or mosquito-to-human transmission event, when a previously-drawn human or mosquito label is assigned to a generated individual. Prior to this point, each mosquito label has been used at most once; and the cardinality of the set of labels of infected humans is necessarily equal to the number of infected humans in the branching process. In particular, the vector of counts $(\mathbf{h}^*, m_1)$ of infected humans and mosquitoes in the branching process and the $P_M^\text{th}$ realisation of the epidemic process is identical until the $\big( \mathcal{M}^{(P_M, r)} - b(\xi^{(0)}) - m(\xi^{(0)}) \big)^\text{th}$ human-to-mosquito or mosquito-to-human transmission event. 

To be more precise, let $Y_r^{P_M*}$ denote the process $Y_r^{P_M}$ with the respective components for uninfected humans and uninfected mosquitoes removed. Mirroring the formulation of \textcite{barbour2010coupling}, let $\mathcal{H}_n$ be the Borel $\sigma$-algebra generated by paths in $\chi^*$ that are initialised in the state $\xi^{(0)}$ and comprise up to $n$ human-to-mosquito and mosquito-to-human transmission events. Then it follows that, for any event $W \in \mathcal{H}_n$,
\begin{align*}
    \Big\{ \omega \in \Omega: &  X_r \in W, \, Y_r^{P_M} \notin W \text{ or }  X_r^{P_M} \notin W, \, Y_r^{P_M} \in W  \Big\} \\
    & \subseteq  \Big\{ \omega = (\omega_1, \omega_2) \in \Omega: \mathcal{M}^{(P_M, r)}(\omega_2) \leq n + b(\xi^{(0)}) + m(\xi^{(0)}) \Big\},
\end{align*}
whereby
\begin{align*}
    \big| \mathbbm{P}[ X_r \in W ] - \mathbbm{P}[ Y_r^{P_M} \in W ] \big| & \leq \EX \big[ \big| \mathbbm{1} \{ X_r \in W \} - \mathbbm{1} \{ Y_r^{P_M} \in W \} \big| \big]\\
    & \leq \EX \big[ \mathbbm{1} \{ X_r \in W, Y_r^{P_M} \notin W \} + \mathbbm{1} \{ X_r \notin W, Y_r^{P_M} \in W \}  \big] \\
    & \leq \mathbbm{P} [ \mathcal{M}^{(P_M, r)}(\omega_2) \leq n + b(\xi^{(0)}) + m(\xi^{(0)}) ] \\
    & \leq \frac{(n+b(\xi^{(0)}) +m(\xi^{(0)}) )^2}{2 P_M} (1 + r).
\end{align*}

This proves part (a).

Following \textcite{ball1995strong}, we can additionally establish strong convergence on the set of sample paths
\begin{align*}
    A = \Big\{ (\omega_1, \omega_2) \in \Omega: \lim_{t \to \infty} X_r(t, \omega_1) = 0 \Big\}
\end{align*}
on which the branching process becomes extinct.

Denote by $T(t, \omega_1)$ the number of human-to-mosquito and mosquito-to-human transmission events in $X_r(\cdot, \omega_1)$ until time $t$. For $\mathbbm{P}_1$-almost all $\omega_1 \in \Omega_1$ such that $\lim_{t \to \infty} X_r(t, \omega_1)=0$, it is necessarily the case that $T(\infty, \omega_1) < \infty$. By construction, there exists $D \in \mathcal{F}_2$ such that $\mathbbm{P}_2(D) = 1$ and $P_M^{-\frac{1}{2}} \mathcal{M}^{(P_M, r)}(\omega_2) \to \mathcal{M}(\omega_2)$ for all $\omega_2 \in D$. We thus consider the set
\begin{align*}
    A^* :=  \Big\{ \omega_1 \in \Omega_1: T(\infty, \omega_1)< \infty  \Big\} \times D,
\end{align*}
for which the symmetric difference with $A$ has zero probability, $\mathbbm{P}(A \triangle A^*) = 0$.

Let $B_{P_M}$ be the set of sample paths on which the epidemic process $Y_r^{P_M}$ and branching process $X_r$ agree for all $t \geq 0$
\begin{align*}
    B_{P_M} : = & \Big\{ (\omega_1, \omega_2) \in \Omega: \sup_{0 \leq t < \infty} \big| Y_r^{P_M}(t, (\omega_1, \omega_2)) - X_r(t, \omega_1) \big|_1 = 0 \Big\},
\end{align*}
and let $C_{P_M}$ be the set of sample paths on which no human or mosquito label has been drawn more than once across all mosquito-to-human and human-to-mosquito transmission events in the branching process
\begin{align*}
     C_{P_M} := \Big\{ \omega = (\omega_1, \omega_2) \in \Omega: \mathcal{M}^{(P_M, r)}(\omega_2) > T(\infty, \omega_1) \Big\}.
\end{align*}

From the argument above, it is necessarily the case that $B_{P_M} \supseteq C_{P_M}$. It thus follows that
\begin{align*}
    A^* \subseteq \bigcup^\infty_{n=1} \bigcap^\infty_{P_M=n} C_{P_M} \subseteq \bigcup^\infty_{n=1} \bigcap^\infty_{P_M=n} B_{P_M}  = \Big\{ \omega \in \Omega: \lim_{P_M \to \infty} \Big( \sup_{0 \leq t < \infty} \big| Y_r^{P_M}(t, \omega) - X_r(t, \omega) \big|_1 \Big) = 0 \Big\}.
\end{align*}

Therefore, in the limit $P_M \to \infty$
\begin{align*}
    \sup_{0 \leq t < \infty} \big| Y_r^{P_M}(t, \omega) - X_r(t, \omega) \big|_1 \to 0 \text{ for } \mathbbm{P}\text{-almost all } \omega \in A.
\end{align*}

\end{proof}

\section{On the probability of disease elimination} \label{sec::disease_extinction}

In Theorem 3.1 of \parencite{mehra2024superinfection}, under the epidemic process $Y^{P_M}_r$, we establish that disease extinction occurs almost surely by coupling the process to a set of independent on-off processes (representing mosquitoes) and infinite server queues with batch arrivals (representing human hosts). Echoing \parencite{ball1983threshold}, the notion of sustained endemic transmission is then formulated in the infinite population limit $P_M \to \infty$ as the infection of infinitely many humans/mosquitoes.

To approximate the probability of disease elimination, as opposed to sustained endemic transmission, when the epidemic process $Y^{P_M}_r$ is initialised with low-level human and mosquito infection, we consider the corresponding probability of global disease extinction under the approximating branching process $X_r$. This is justified by Theorem \ref{theorem::superinf_branching_approx} where we established that there is a total variation bound, of order $O(P_M^{2 \kappa - 1})$, which holds until $o(P_M^\kappa)$ human-to-mosquito or mosquito-to-human transmission events have occurred (where $\kappa < 1/2$ is an arbitrary constant), for the branching process $X_r$ relative to the epidemic process $Y_r^{P_M}$.

A key determinant of disease extinction under the approximating branching process $X_r$ is the expected number of additional mosquito inocula attributable to a given mosquito inoculum after one cycle of human-to-mosquito-to-human transmission. Following reasoning analogous to that of \textcite{mehra2024superinfection}, we can show that this is given by
\begin{align}
    R_0^2 = \overbrace{\underbrace{\beta p \vphantom{\int^\infty_0 \bigg( 1 - \frac{1-e^{-\gamma t}}{ 1 + \nu \big[ 1 - \frac{\eta}{\eta - \gamma} (e^{-\gamma t} - e^{-\eta t}) \big]} \bigg) d t} }_{\substack{\text{transmission} \\ \text{rate per infected} \\ \text{mosquito}}} \; \underbrace{\frac{1}{g} \vphantom{\int^\infty_0 \bigg( 1 - \frac{1-e^{-\gamma t}}{ 1 + \nu \big[ 1 - \frac{\eta}{\eta - \gamma} (e^{-\gamma t} - e^{-\eta t}) \big]} \bigg) d t} }_{\substack{\text{expected} \\ \text{mosquito} \\ \text{lifetime}}}}^\text{mosquito-to-human cycle} \; \cdot \; \overbrace{\underbrace{\beta q r \vphantom{\int^\infty_0 \bigg( 1 - \frac{1-e^{-\gamma t}}{ 1 + \nu \big[ 1 - \frac{\eta}{\eta - \gamma} (e^{-\gamma t} - e^{-\eta t}) \big]} \bigg) d t} }_{\substack{\text{transmission} \\ \text{rate per blood} \\ \text{infected human}}} \; \underbrace{\int^\infty_0 \bigg( 1 - \frac{1-e^{-\gamma t}}{ 1 + \nu \big[ 1 - \frac{\eta}{\eta - \gamma} (e^{-\gamma t} - e^{-\eta t}) \big]} \bigg) d t}_{\substack{\text{expected cumulative duration of blood-stage infection} \\ \text{attributable to a single mosquito inoculum} \\ \text{(adjusted for within-inoculum superinfection)}}}}^\text{human-to-mosquito cycle}. \label{eq:R0}
\end{align}
In Theorem \ref{theorem::branching_process_extinction} below, we explicitly characterise the probability of global disease extinction under the approximating branching process $X_r$, concluding that global extinction occurs almost surely in the case $R_0<1$ while there is a non-zero probability of survival in the case $R_0>1$.

\textbf{\emph{Remark}}: The functional law of large numbers for the epidemic process $Y_r^{P_M}$, which constitutes a deterministic compartment model, exhibits an analogous threshold phenomenon \parencite{mehra2024superinfection}. In the infinite population limit $P_M \to \infty$, we show in Theorem 3.2 of \parencite{mehra2024superinfection} that the sample paths of $Y^{P_M}_r$ converge to the solution of an infinite-dimensional system of ODES, which can subsequently be reduced into an integrodifferential equation (IDE) governing the prevalence of mosquito infection \parencite{mehra2024superinfection}. The parameter $R_0^2$ defined in Equation (\ref{eq:R0}) is equal to the reproduction number of this deterministic system (Theorem 4.1 of \parencite{mehra2024superinfection}): in the case $R_0<1$, we show that the disease-free equilibrium is uniformly asymptotically stable (for the reduced IDE) \parencite{brauer1978asymptotic}, and that no endemic equilibrium exists; in the case $R_0>1$, we prove the existence of a unique endemic equilibrium and provide a sufficient condition for it to be uniformly asymptotically stable (for the reduced IDE).


\begin{theorem}\quad \label{theorem::branching_process_extinction}
\begin{itemize}
    \item[(a)] If $R_0 \leq 1$ and $|X_r(0)|_1 < \infty$, then $\lim_{t \to \infty} |X_r(t)|_1 = 0$ almost surely, that is, global disease extinction occurs almost surely under the approximating branching process.
    \item[(b)] If $R_0 > 1$, let $q_{i, j}$ and $q_m$ denote the probability of global disease extinction when the process  $X_r$ is initialised with a single particle of type $H_{i, j}$ or $M$ respectively. Then $q_{i,j} \in (0, 1)$ for $(i,j) \in \mathbbm{Z}^2_{\geq 0} \setminus \{ (0, 0) \}$ and $q_m \in (0, 1)$ are given by the unique solution to the system
    \begin{align}
    \begin{split}
        & q_{0,j} = \prod^j_{i=1} \frac{\gamma i}{\gamma i + \beta q r ( 1 - q_m)}, \qquad j \in \mathbbm{N} \\
        & q_{i, 1} = q_{i+1, 0} \qquad \qquad \qquad \qquad \quad i \in \mathbbm{Z}_{\geq 0} \\
        & q_{i, j} = \frac{\gamma j _{i, j-1} + \eta i q_{i-1, j+1}}{\eta_i + \gamma j + \beta q r (1-q_m)}  \quad \, \, \, (i, j) \in \mathbbm{N}^2 \\
        & \sum^\infty_{i=0} \frac{1}{\nu + 1} \Big( \frac{\nu}{\nu + 1} \Big)^{i} q_{i, 1} = 1 + \frac{g }{\beta p} \Big( 1 - \frac{1}{q_m} \Big)
    \end{split} \label{eq:extinction_prob_supercritical}
    \end{align}
    in the domain $q_{i, j} \in (0, 1)$ and $q_m \in (0, 1)$.
\end{itemize}
\end{theorem} 

\begin{proof}
Denote by
\begin{align*}
    q_{i, j} &= \mathbbm{P} \Big[ \lim_{t \to \infty} \big| X_r(t) \big|_1 = 0 \, \big| \, X_r(0) = (\mathbf{e_{i,j}}, 0) \Big] \\
    q_{m} &= \mathbbm{P} \Big[ \lim_{t \to \infty} \big| X_r(t) \big|_1 = 0 \, \big| \, X_r(0) = (\mathbf{0}, 1) \Big]
\end{align*}
the probability of global extinction when the approximating branching process is initialised by a single particle of type $H_{i, j}$ or $M$ respectively.

To characterise global extinction for the continuous-time process $X_r(t)$, it suffices to consider the embedded infinite-dimensional Galton-Watson branching process. This is specified by the offspring distribution for each particle type aggregated over its `lifetime'. Here, we interpret the lifetime to govern the time until which the particle is either removed from the system; transforms to another type; or leads to onward transmission, whereby it `dies' but immediately concurrently generates a particle of the same type. We can use the transition rates (\ref{eq::human_clearance_branching_1}) to (\ref{eq::h_to_m_branching}) to construct transition probabilities for the states of the process. Set
\begin{align*}
    \mathbf{z} := ( z_m, z_{0, 1}, z_{1, 0}, z_{0, 2}, z_{1, 1}, z_{2, 0}, \dots) \in [0, 1]^{\mathbbm{N}}.
\end{align*}
Then we can formulate infinite-dimensional probability generating functions for the offspring distribution for type $H_{i, j}$ particles
\begin{align}
    G_{i,j}(\mathbf{z}) = \begin{cases}
        \frac{\beta q r}{\beta q r + \gamma j + \eta i} z_{i, j} z_m + \frac{\gamma j}{\beta q r + \gamma j + \eta i} z_{i, j-1} + \frac{\eta i}{\beta q r + \gamma j + \eta i} z_{i-1, j+1} &  (i, j) \in \mathbbm{N}^2 \\
        \frac{\beta q r}{\beta q r + \gamma j } z_{0, j} z_m + \frac{\gamma j}{\beta q r + \gamma j} z_{0, j-1}  & (i, j) \in \{ 0 \} \times \mathbbm{Z}_{\geq 2}  \\
        \frac{\beta q r}{\beta q r + \gamma } z_{0, j} z_m + \frac{\gamma}{\beta q r + \gamma}  &  (i, j) = (0, 1) \\
        z_{i-1, 1} &  (i, j) \in \mathbbm{N} \times \{0 \}
    \end{cases} \label{eq:human_offspring_pgf}
\end{align}
and type $M$ particles
\begin{align}
    G_m(\mathbf{z}) = \frac{g}{g + \beta p}  + \frac{\beta p}{g + \beta p} z_m \sum^\infty_{i=0} \frac{1}{\nu + 1} \Big( \frac{\nu}{\nu + 1} \Big)^i z_{i, 1} \label{eq:mosquito_offspring_pgf}
\end{align}
respectively. Define
\begin{align*}
    \mathbf{G}(\mathbf{z}) = \big(  G_m(\mathbf{z}),  G_{0,1}(\mathbf{z}),  G_{1,0}(\mathbf{z}),  G_{0,2}(\mathbf{z}),  G_{1,1}(\mathbf{z}),  G_{2,0}(\mathbf{z}) \dots).
\end{align*}

Since the process is irreducible and possesses countably infinitely-many types, it follows from \textcite{bertacchi2020extinction} that the global extinction probability vector $\mathbf{q}$ is the componentwise minimal element of the fixed point set
\begin{align*}
    S = \{ \mathbf{s} \in [0, 1]^\mathbbm{N}: G(\mathbf{s}) = \mathbf{s} \big\}.
\end{align*}

Setting $s_{i, j} = G_{i, j}(\mathbf{s})$ for each $(i, j) \in \mathbbm{Z}_{\geq 0}^2 \setminus \{ (0, 0) \}$, we can use Equation (\ref{eq:human_offspring_pgf}) to show that any fixed point $\mathbf{s} \in S$ must satisfy
\begin{align}
\begin{split}
    s_{0,j} &=  \prod^j_{i=1} \frac{\gamma i}{\gamma i + \beta q r ( 1 - s_m)}, \qquad j \in \mathbbm{N} \\
    s_{i, 1} &= s_{i+1, 0} \qquad \qquad \qquad \qquad \quad i \in \mathbbm{Z}_{\geq 0} \\
    s_{i, j} &= \frac{\gamma j s_{i, j-1} + \eta i s_{i-1, j+1}}{\eta_i + \gamma j + \beta q r (1-s_m)}  \quad \, \, \, (i, j) \in \mathbbm{N}^2.
\end{split} \label{eq::branching_process_fixed_points}
\end{align}
This system of equations can be evaluated along the upper triangle (that is, for each $i=0,1,\dots, n$, iterating over $j=0, \dots, n-i$ for increasing $n$) to yield an expression for each $s_{i,j}$ as a function of $s_m$ only. We can show by induction that, for each $(i, j) \in \mathbbm{Z}_{\geq 0}^2 \setminus \{ (0, 0)\}$, as a function of $s_m$ on the domain $[0, 1]$, $s_{i,j}$ is non-negative, monotonically increasing and convex, with $s_{i, j} \in (0, 1)$ for all $s_m \in [0, 1)$. From the offspring PGF $G_m$, we note that $s_m$ must satisfy
\begin{align}
    f_1(s_m) := \sum^\infty_{i=0} \frac{1}{\nu + 1} \Big( \frac{\nu}{\nu + 1} \Big)^i s_{i,1}(s_m) = 1 + \frac{g}{\beta p} - \frac{g}{\beta p} \frac{1}{s_m} := f_2(s_m). \label{eq::branching_process_fixed_points_2}
\end{align}
The series that defines $f_1(s_m)$ is uniformly convergent, and so we can exchange the order of summation and differentiation to verify that $f_1(s_m)$ is likewise a non-negative, monotonically increasing and convex function of $s_m$ on the domain $[0, 1]$, with $f_1(1) = 1$. In contrast, $f_2(s_m)$ is a monotonically increasing, concave function of $s_m$, with $f_2(s_m) \to -\infty$ in the limit $s_m \to 0$. Noting that $f_1(1) = f_2(1) = 1$, we can use the simple geometric argument detailed in \textcite{mehra2024superinfection} to recover two cases:
\begin{itemize}
    \item If $f_1'(1) \leq f_2'(1)$, then $f_1(s_m) \neq f_2(s_m)$ for all $s_m \in [0, 1)$, whereby $S = \{ \mathbf{1} \}$
    \item If $f_1'(1) > f_2'(1)$, then there exists a unique solution $q_m \in (0, 1)$ such that $f_1(q_m) = f_2(q_m)$, whereby $S = \{ \mathbf{1}, \mathbf{q} \}$ where for each $(i, j) \in \mathbbm{Z}_{\geq 0}^2 \setminus \{ (0, 0)\}$, we recover $q_{i,j}$ by plugging $s_m = q_m$ into Equation (\ref{eq::branching_process_fixed_points}), and it is necessarily the case that each $q_{i, j} \in (0, 1)$.
\end{itemize}

It remains to calculate $f_1'(1)$. Treating each $s_{i,j}$, $(i, j) \in \mathbbm{Z}_{\geq 0}^2 \setminus \{ (0, 0)\}$ as a function of $s_m$, the equality $s_{i, j} = G_{i,j}(\mathbf{s})$ implies that
\begin{align}
  \frac{d s_{i,j}}{d s_m} \bigg|_{s_m = 1} = \begin{cases}
        \frac{\beta q r}{\gamma j + \eta i} + \frac{\gamma j}{\gamma j + \eta i} \frac{d s_{i,j-1}}{d s_m} \big|_{s_m = 1} + \frac{\eta i}{\gamma j + \eta i} \frac{d s_{i-1,j+1}}{d s_m} \big|_{s_m = 1} & (i, j) \in \mathbbm{N}^2 \\
        \frac{\beta q r}{\gamma j} + \frac{d s_{0,j-1}}{d s_m} \big|_{s_m = 1} & (i, j) \in \{ 0 \} \times \mathbbm{N}  \\
        \frac{d s_{i-1, 1}}{ds_m} \big|_{s_m = 1} & (i, j) \in  \mathbbm{N} \times  \{ 0 \}.
    \end{cases}\label{eq::branching_process_h_to_m_mean}
\end{align}
From the structure of Equation (\ref{eq::branching_process_h_to_m_mean}), it is immediate that $\frac{d s_{i,j}}{d s_m} \big|_{s_m=1}$ can be interpreted as the expected number of human-to-mosquito transmission events until a human with hypnozoite burden $i$ and MOB $j$ completely clears infection, assuming that human-to-mosquito transmission occurs at constant rate $\beta q r$ during periods of blood-stage infection and the individual is not subject to further parasite inoculation. The activation times of different hypnozoites, and the clearance times of different blood-stage infections are independent. A blood-stage infection present at time zero is still circulating in the bloodstream at time $t$ with probability $e^{-\gamma t}$, while a hypnozoite present in the liver at time zero contributes to blood-stage infection at time $t$ with probability $\frac{\eta}{\eta - \gamma} ( e^{-\gamma t} - e^{-\eta t})$ \parencite{mehra2022hypnozoite}. Therefore, we can write
\begin{align*}
    \frac{d s_{i,j}}{d s_m} \bigg|_{s_m=1} = \int^\infty_{0} \bigg( 1 - (1-e^{-\gamma t})^j \bigg[ 1 - \frac{\eta}{\eta - \gamma} (e^{-\gamma t} - e^{-\eta t}) \bigg]^i \bigg) d t.
\end{align*}
It then follows that
\begin{align*}
    f_1'(1) &= \beta q r \sum^\infty_{i=0} \frac{1}{\nu + 1} \Big( \frac{\nu}{1 + \nu} \Big)^i \cdot \frac{d s_{i,1}}{d s_m} \bigg|_{s_m=1} = \beta q r \int^\infty_0 \bigg( 1 - \frac{1-e^{-\gamma t}}{1 + \nu \big[ 1 - \frac{\eta}{\eta - \gamma}(e^{-\gamma t} - e^{-\eta t}) \big]} \bigg) d t
\end{align*}
where interchanging the order of summation and integration is justified by the Fubini-Tonelli theorem since the integrand is non-negative. As noted in \textcite{mehra2024superinfection}, we can interpret the quantity $T_{m \to h} := f'(1)/(\beta q r)$ as the expected duration of blood-stage (super)infection for each mosquito inoculum. Denote by
\begin{align*}
    R_0^2 := f_1'(1)/f_2'(1) = \frac{\beta^2 p q r T_{m \to h}}{g}.
\end{align*}

Then we obtain the result:
\begin{itemize}
    \item In the case $R_0 \leq 1$, $\lim_{t \to \infty} | X_r(t)|_1 = 0$ almost surely.
    \item In the case $R_0 > 1$, the global disease extinction probabilities $q_m \in (0, 1)$ and $q_{i, j} \in (0, 1)$, $(i, j) \in \mathbbm{Z}_{\geq 0}^2 \setminus \{(0, 0)\}$ are given by the unique solution to Equations (\ref{eq::branching_process_fixed_points}) and (\ref{eq::branching_process_fixed_points_2}) in this domain.
\end{itemize}
\end{proof}

\section{Illustrative results} \label{sec::illustrative_results}

We now apply the branching process approximation $X_r$ to characterise the probability of disease elimination in scenarios of epidemiological interest, where the epidemic process $Y_r^{P_M}$ is initialised with a low-level burden of infection in the human and mosquito population. Assumed parameter values (Table \ref{tab:param_summary}) have been chosen to reflect the observation that asymptomatic parasitemias, which constitute the bulk of the \textit{P. vivax} blood-stage infection burden in endemic settings, persist in the bloodstream for extended periods but oscillate to submicroscopic, non-transmissible densities \parencite{nguyen2018persistence}. Accordingly, we accommodate a relatively slow rate of clearance $\gamma=1/60$ day$^{-1}$ per blood-stage brood \parencite{white2014modelling}, but enforce a low probability of successful human-to-mosquito transmission $q=0.04$ when a mosquito takes a bloodmeal from a human with MOB$\geq$1. We vary the ratio $r$ of the mosquito to human population size to reflect transmission settings where individuals are subject to varying degrees of mosquito exposure.

\begin{table}[h]
    \centering
    \begin{tabular}{|c|p{8.4cm}|c|c|}
        \hline
        \textbf{Parameter} & \textbf{Interpretation} & \textbf{Value} & \textbf{Source}  \\ \hline \hline
        $\eta$ & Hypnozoite activation rate & 1/170 day$^{-1}$ & \parencite{Mehra2024b} \\ \hline 
        $\nu$ & Average number of hypnozoites (per mosquito inoculum) that are destined to activate & 2.7 & \parencite{Mehra2024b} \\ \hline 
        $\gamma$ & Rate of clearance for each blood-stage brood & 1/60 day$^{-1}$ & \parencite{white2014modelling}$^*$  \\ \hline
        $\beta$ & Rate at which each mosquito seeks human bloodmeals & 0.25 day$^{-1}$ & \parencite{garrett1969malaria} \\  \hline
        $g$ & Rate of mosquito death/replacement by uninfected progeny  & $0.1$ day$^{-1}$ & \parencite{fraser2021estimating} \\ \hline
        $p$ & Probability of mosquito-to-human transmission when an infected mosquito bites a human host & 0.55 & \parencite{smith2010quantitative} \\ \hline
        $q$ & Probability of human-to-mosquito transmission when an uninfected mosquito takes a bloodmeal from a human with MOB$\geq1$ & 0.04 & assumed \\ \hline
        $r$ & Ratio of mosquito to human population size & [0.88, 4] & varied \\ \hline
    \end{tabular}
    \caption{Summary of model parameters, their interpretations and assumed values. $^*$Secondary source.}
    \label{tab:param_summary}
\end{table}

\subsection{Probability of re-instated transmission}


Imported cases of \textit{P. vivax} from neighbouring endemic regions can pose a key challenge in near-elimination and post-elimination settings
\parencite{cotter2013changing, 
yin2023imported, wangdi2024estimating}. Here, we characterise the probability of re-instated endemic transmission when an infected individual enters a setting where \textit{P. vivax} malaria has previously been eliminated, but which still harbours malaria-transmitting vectors. For simplicity, we assume that the infected individual has recently been bitten and therefore harbours a primary blood-stage infection and a geometrically-distributed hypnozoite burden (with mean size $\nu$) at the time of entry. We thus approximate the probability of re-instated transmission by
\begin{align}
    \mathbbm{P}(\text{re-instated transmission}) \approx 1 - \sum^\infty_{i=0} \frac{1}{\nu+1} \Big( \frac{\nu}{\nu+1} \Big)^i q_{i, 1} = \frac{g}{\beta p} \bigg( \frac{1}{q_m} -1 \bigg) \label{eq::reinstated_transmission}
\end{align}
where $q_{i,j}$ and $q_m$ denote the respective probabilities of global disease extinction under the branching process approximation $X_r$ initialised with a single particle of type $H_{i,j}$ or $M$ respectively (Theorem \ref{theorem::branching_process_extinction}).

\begin{figure}[h]
    \centering
    \includegraphics[width=0.6\linewidth]{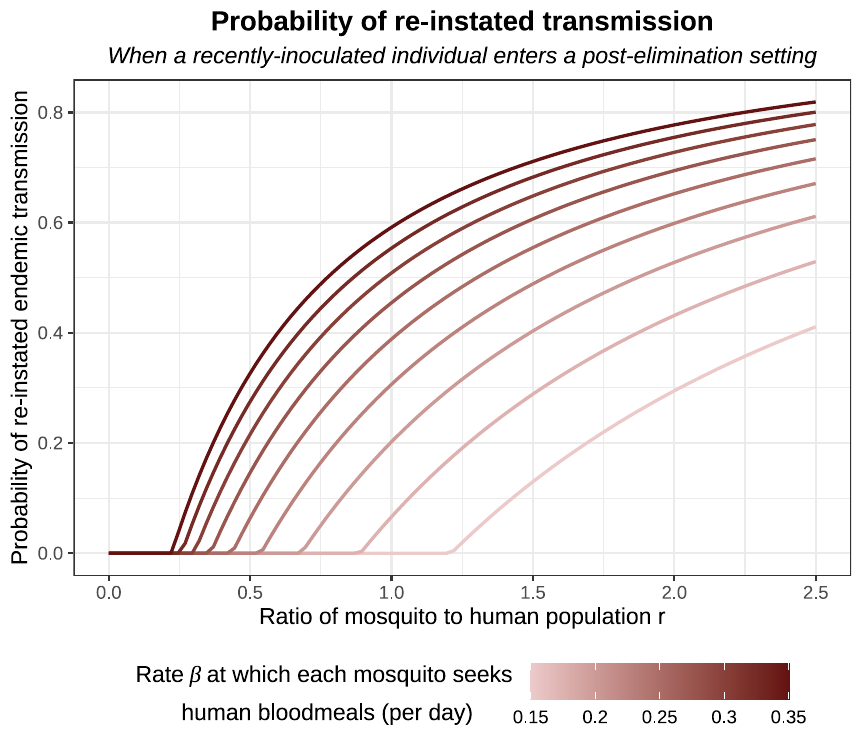}
    \caption{Probability of re-instated endemic transmission (\ref{eq::reinstated_transmission}) when a recently inoculated individual (with a primary infection and geometrically-distributed hypnozoite burden of mean size $\nu$) enters a post-elimination setting, as a function of two key entomological parameters: the ratio $r$ of the mosquito to human population, and the rate $\beta$ at which each mosquito seeks human bloodmeals. All other parameter values are detailed in Table \ref{tab:param_summary}.}
    \label{fig:reintroduction}
\end{figure}

We find that the probability of re-instated transmission (\ref{eq::reinstated_transmission}) is heavily-dependent on the availability of mosquito vectors, manifest in the ratio of the mosquito to human population size $r$ and the rate $\beta$ at which each mosquito seeks human blood meals (Figure \ref{fig:reintroduction}). These entomological parameters may vary broadly across endemic settings, due to mosquito feeding preferences (zoophagy vs anthrophagy) \parencite{garrett1964human}, mosquito species composition \parencite{van2023anopheles, fansiri2024species}, and environmental factors like land cover which can modulate mosquito abundance \parencite{hernandez2020effect}. Seasonality (which we have not modelled) is another important determinant of vector populations \parencite{fansiri2024species}. Nonetheless, our results suggest substantial heterogeneity in the risk of re-introduction across post-elimination settings, and reinforce the need to monitor imported \textit{P. vivax} cases. 

\subsection{Probability of elimination following (idealised) mass drug administration}

We additionally consider the probability of elimination following a mass drug administration (MDA) intervention, whereby radical curative treatment (which targets both blood-stage broods and the hypnozoite reservoir) is widely administered across the human population. We consider a hypothetical setting with perfect population-wide coverage, and assume that MDA clears all pre-existing blood-stage broods. We further assume that an extended period of prophylactic protection against blood-stage infection in the human population interrupts transmission for a sufficiently long period to ensure all pre-existing infected mosquitoes are replaced by uninfected progeny. However, we allow for sustained hypnozoite carriage in spite of MDA, with each hypnozoite surviving radical curative treatment with probability $p_\text{rad}$ \parencite{mehra2022hypnozoite}. This represents a highly-idealised scenario: in practice, population-wide coverage of radical curative treatment is constrained by the prevalence of glucose-6-phosphate dehydrogenase (G6PD) deficiency (mean 8\% in malaria endemic settings), which poses a risk of life-threatening haemolysis following radical curative treatment \parencite{baird2018primaquine}; perfect coverage in the G6PD-normal population is operationally difficult to achieve; and blood-stage efficacy and prophylaxis may be imperfect.

We approximate the pre-MDA hypnozoite burden in the population by the endemic equilibrium solution to the deterministic compartment model which arises as the functional law of large numbers for the epidemic process $Y_r^{P_M}$ \parencite{mehra2024superinfection}. Given $R_0>1$, the distribution of the hypnozoite burden at the endemic equilibrium solution is negative binomial with mean $\beta p r I_M^* \nu/\eta$ and shape parameter $\beta p r I_M^*/\eta$, where $I_M^* \in (0, 1]$ is given by the (unique) non-zero solution to equation
\begin{align}
    \frac{g I^*_M}{\beta q (1-I^*_M )} = 1 - e^{-\frac{\beta q R_0^2}{g} I_M^*} \label{ide_im_ss}.
\end{align}

For this endemic equilibrium to be uniformly asymptotically stable (in the sense of \parencite{brauer1978asymptotic} for the reduced IDE), it is sufficient that 
\begin{align}
    I_M^* > \frac{1 + \frac{\beta q}{\beta q+g} - \sqrt{\big( 1 - \frac{\beta q}{\beta q+g} \big)^2 + 4 \frac{\beta q}{\beta q+g} \frac{1}{R_0^2} }}{2} \label{eq:stable_endemic_eq}
\end{align}
(Theorem 4.1 of \parencite{mehra2024superinfection}); we verify that this holds for all parameter combinations considered in this paper. Given a per-hypnozoite survival probability of $p_\text{rad}$, the population frequency distribution for the post-MDA hypnozoite burden (used to approximate the within-host post-MDA burden) is likewise negative binomial, but with mean $\beta p r I_M^* \nu p_\text{rad}/\eta$ and shape parameter $\beta p r I_M^*/\eta$.

For a closed human population of size $P_H = r P_M$, the probability of disease elimination following MDA is therefore approximated to be
\begin{align}
    \mathbbm{P}(\text{elimination after idealised MDA}) \approx \Bigg[ \sum^\infty_{i=0} q_{i, 0} \frac{\Gamma(i + \beta p r I_M^*/\eta) }{i! \cdot \Gamma(\beta p r I_M^*/\eta)} \frac{(\beta p r I_M^* \nu p_\text{rad})^i \cdot \eta^{\beta p r I_M^* /\eta } }{(\beta p r I_M^* \nu p_\text{rad} + \eta)^{i + \beta p r I_M^*/\eta}}  \Bigg]^{P_H} \label{eq:MDA_elimination}
\end{align}
where $q_{i,0}$ denotes the probability of global disease extinction under the branching process approximation $X_r$ initialised with a single particle of type $H_{i,0}$ (Theorem \ref{theorem::branching_process_extinction}).

\begin{figure}[h]
    \centering
    \includegraphics[width=0.6\linewidth]{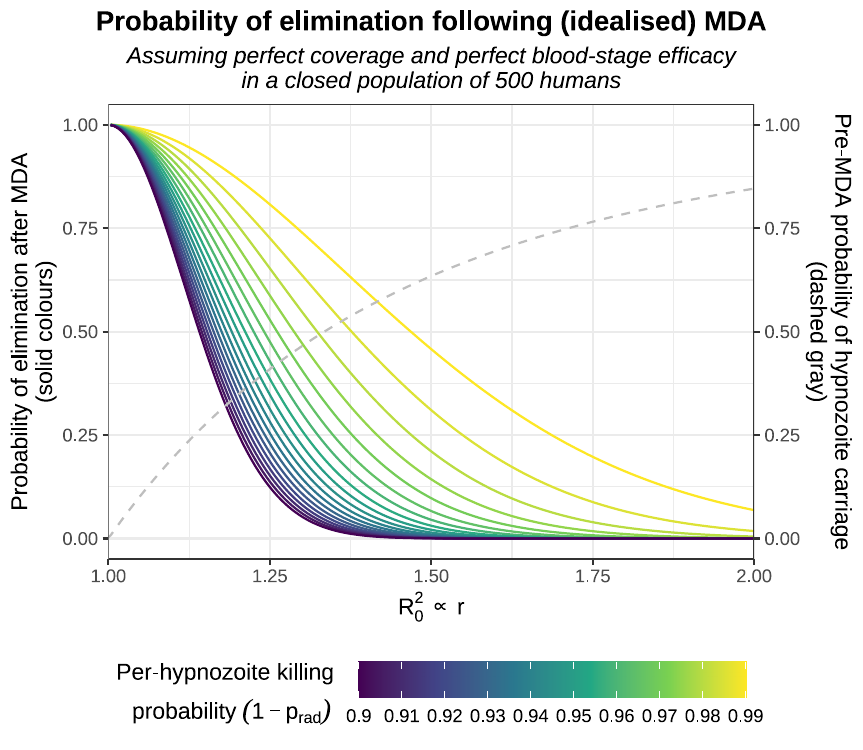}
    \caption{Probability of disease elimination (\ref{eq:MDA_elimination}) following an idealised MDA intervention in a population of 500 people, as a function of the per-hypnozoite killing probability $(1-p_\text{rad})$ and the reproduction number $R_0^2$ (\ref{eq:R0}) when the ratio of the mosquito to human population $r$ is varied (whereby $R_0^2 \propto r$). The gray dashed line shows the pre-MDA probability of hypnozoite carriage $1-(1 + \nu)^{-\beta p r I_M^* \nu / \eta}$ based on the endemic equilibrium solution to the functional law of large numbers for the epidemic process $Y_r^{P_M}$, verified to be be uniformly asymptotically stable using condition  (\ref{eq:stable_endemic_eq}) \parencite{mehra2024superinfection}. All other parameter values are detailed in Table \ref{tab:param_summary}.}
    \label{fig:MDA_elimination}
\end{figure}

The probability of disease elimination following a hypothetical highly-idealised MDA intervention in a closed population of $P_H=500$ people (Equation (\ref{eq:MDA_elimination})) is shown in Figure \ref{fig:MDA_elimination} as a function of the per-hypnozoite killing probability $(1-p_\text{rad})$ and the reproduction number $R_0^2$ (\ref{eq:R0}) when the ratio of the mosquito to human population $r$ is varied (whereby $R_0^2 \propto r$); overlaid in gray is the probability of hypnozoite carriage prior to MDA. Even under a highly-idealised scenario, we find that disease elimination using MDA is probable only in a relatively narrow range of transmission settings, where a substantial proportion of individuals is predicted to be hypnozoite-free at baseline. In such settings, serological testing to identify potential hypnozoite carriers may aid in mitigating overtreatment \parencite{obadia2022developing, Mehra2024b}. 

The per-hypnozoite killing probability $(1-p_\text{rad})$ is a key determinant of the predicted probability of disease elimination. Multiple rounds of MDA may augment the probability of disease elimination \parencite{anwar2024investigation}, assuming a multiplicative effect on hypnozoite survival whereby the per-hypnozoite probability of survival following $n$ rounds of MDA is $p_\text{rad}^n$ \parencite{mehra2022hypnozoite}. Accounting for multiple rounds of MDA, however, requires modelling the replenishment of the hypnozoite reservoir between successive MDA rounds \parencite{anwar2024investigation}, which is beyond the present scope.

\section{Discussion} \label{sec::discussions}
Hypnozoite dynamics are an important determinant of the epidemiology of \textit{P. vivax} malaria, posing significant challenges for control and elimination efforts. Superinfection with blood-stage broods derived from different primary infections and hypnzoite activation events is a key consideration in within-host models of \textit{P. vivax}. We have previously proposed a stochastic epidemic process $Y_r^{P_M}$, formulated as a Markov population process with countably many types, to describe \textit{P. vivax} dynamics in a closed population whilst adjusting for hypnozoite accrual and superinfection \parencite{mehra2024superinfection}. Here, we construct a Markovian branching process with countably many types $X_r$ to approximate the early stages of the epidemic process $Y_r^{P_M}$, intuiting that the prevalence of both mosquito infection, and host superinfections derived from multiple mosquito inocula, may be negligibly low in the early stages of an epidemic. We adopt the classical coupling argument of \textcite{ball1995strong} to establish a total variation bound, of order $O(P_M^{2 \kappa -1})$ in the infinite population limit $P_M \to \infty$ (with the ratio $r$ of the mosquito to human population held fixed), that is valid until $o(P_M^{\kappa})$ human-to-mosquito or mosquito-to-human transmission events have occurred (where $\kappa < 1/2$ is an arbitrary constant) (Theorem \ref{theorem::superinf_branching_approx}a); on the set of sample paths which culminate in disease extinction for the branching process $X_r$, we additionally establish strong convergence in the limit $P_M \to \infty$ (Theorem \ref{theorem::superinf_branching_approx}b). We characterise the probability of global disease extinction for the branching process $X_r$ (Theorem \ref{theorem::branching_process_extinction}) to approximate the probability of disease elimination, as opposed to sustained endemic transmission (defined as infinite human/mosquito infection in the infinite population limit $P_M \to \infty$ \parencite{ball1983threshold}), for the epidemic process $Y_r^{P_M}$. In particular, we establish a threshold phenomenon analogous to the functional law of large numbers for the epidemic process $Y_r^{P_M}$ \parencite{mehra2024superinfection}, proving that global disease extinction occurs almost surely under the branching process $X_r$ given the reproduction number $R_0^2<1$ (Equation (\ref{eq:R0})). We then apply the branching process approximation to several illustrative scenarios of epidemiological interest. 


The questions which motivated the present body of work --- namely, the prospect of disease elimination following targeted control and the risk of re-instated transmission due to imported cases --- are of fundamental importance in near-elimination and post-elimination settings. These questions have been explored extensively in the modelling literature for malaria. For non-relapsing \textit{P. falciparum} malaria, \textcite{cohen2010absolute, pemberton2017stochastic} construct simple single-type branching processes to characterise the probability of disease elimination following importation events and MDA respectively, assuming either Poisson or negative binomial distributions for the number of secondary \textit{P. falciparum} cases attributable to each blood-stage infection. \textcite{mbogo2018stochastic} instead propose a continuous-time Markovian process with finitely-many types, accounting for infected mosquitoes, and infected vs `pseudo-recovered' humans, to likewise characterise disease extinction for non-relapsing \textit{P. falciparum}. However, to the best of our knowledge, branching process models have not previously been used to characterise the probability of disease extinction for \textit{P. vivax}, with additional complexity attributable to the hypnozoite reservoir. Importantly, the inoculation of (unbounded) hypnozoite batches necessitates the construction of a branching process with countably infinitely-many types, to avoid double-counting the contribution of within-batch superinfection which can appear early on in the epidemic process.

The dynamics of disease elimination for \textit{P. vivax} have instead been addressed by simulating approximate stochastic trajectories under several model frameworks. \textcite{anwar2024investigation} simulate approximate sample paths for a multiscale model predicated on the within-host framework of \parencite{mehra2022hypnozoite}, that is consequently adjusted for hypnozoite accrual and superinfection, using a $\tau$-leaping algorithm to interrogate the probability of \textit{P. vivax} elimination following multiple rounds of MDA. \textcite{champagne2024quantifying} construct a compartment model which dichotomises hypnozoite carriage but accounts for a broad range of interventions, and likewise employ $\tau$-leaping to accommodate extinction in a stochastic framework. \textcite{roy2013potential} simulate a human compartment model with Erlang-distributed inter-relapse intervals, with mosquito inoculation modelled implicitly through a seasonal force of infection that incorporates stochastic `environmental noise'. Previously developed spatiotemporal agent-based models \parencite{pizzitutti2015validated, gharakhanlou2019developing}, which are predicated on the binarisation of hypnozoite carriage but capture fine-scale heterogeneity, arguably also offer natural simulation frameworks to examine disease elimination. The branching process approximation derived in the present paper, in contrast, yields a rigorous means to interrogate the probability of elimination for \textit{P. vivax} without relying on simulation, whilst accommodating both hypnozoite accrual and superinfection. However, it is predicated on a number of simplifying assumptions, including the absence of seasonal forcing and routine treatment, which are not made in some or all of these simulation-based frameworks.

In summary, by coupling a Markovian branching process with countably infinitely-many types to a previously-developed stochastic epidemic model accounting for both hypnozoite accrual and superinfection \parencite{mehra2024superinfection}, we interrogate the probability of disease elimination for \textit{P. vivax} malaria in epidemiologically relevant contexts. The underlying principle of accommodating within-batch infection dynamics applies in generality when constructing branching process approximations for epidemic models of superinfection (predicated on the simplifying assumption of independence between broods). The coupling argument of \parencite{ball1995strong}, for instance, can be used to establish a two-type Markovian branching process approximation for the classical model of superinfection for non-relapsing \textit{P. falciparum} \parencite{bailey1957}, in which each mosquito-to-human transmission event establishes a single (independent) blood-stage brood. An analogous argument can also be applied to show that epidemic models with concomitant immunity, whereby pre-existing infection inhibits further infection \parencite{barbour2010coupling}, and epidemic models with superinfection, with the independent co-existence of broods derived from different transmission events, share an approximating branching process given identical within-batch dynamics.




\section*{Acknowledgements}
SM gratefully acknowledges a Keble College Sloane Robinson/Clarendon Scholarship and a Nuffield Department of Clinical Medicine Studentship from the University of Oxford. This work was supported by the Australian Research Council Centre of Excellence for the Mathematical Analysis of Cellular Systems (project number CE230100001).

\section*{Data Availability Statement}
Data sharing is not applicable to this article as no datasets were generated or analysed during the current study.

\section*{Competing Interests}
The authors have no competing interests to declare that are relevant to the content of this article.

\printbibliography

\end{document}